\documentstyle[emulateapj]{article}
\begin{document}

\title{A NEW INTERPRETATION OF CHAIN GALAXIES AT HIGH REDSHIFT}

\author{Yoshiaki Taniguchi, \& Yasuhiro Shioya}

\affil{Astronomical Institute, Graduate School of Science, 
       Tohoku University, Aramaki, Aoba, Sendai 980-8578, Japan}

\begin{abstract}
We propose a possible new model for the formation of
chain galaxies at high redshift. Our model is summarized as follows.
Step I: Successive merging of subgalactic gas clumps results in
the formation of a galaxy with a mass of $10^{11-12} M_\odot$
at redshift $z \sim 5$.
Step II: Subsequently, supernova explosions occur inside the galaxy
and then blow out as a galactic wind (or a superwind).
This wind expands into the intergalactic space and then causes a
large-scale shell with a radius of several hundreds of kpc.
Since this radius may be smaller than the typical separation between
galaxies, interactions of shells may also occur, resulting in
the formation of a large-scale gaseous slab. 
Step III: Since the shell or the slab can be regarded as a gaseous sheet, 
filament-like gravitational instability is expected to occur.
Step IV: Further gravitational instability occurs in each filament,
leading to intense star formation along the filament. This is the
chain galaxy phase.
Step V: The filament collapses gravitationally into one spheroidal
system like an elliptical galaxy within one dynamical timescale of
the filament ($\sim 10^8$ yr). Therefore, it seems quite difficult
to find remnants of chain galaxies. 

We also discuss that shocked shells driven by superwinds may be 
responsible for some Lyman limit systems and damped Ly$\alpha$
systems because their H {\sc i} column densities are expected to
be $N_{\rm HI} \gtrsim 10^{19}$ cm$^{-2}$.
\end{abstract} 


\section{INTRODUCTION}

Recent high-resolution optical imaging studies of some deep fields
have revealed a new population of galaxies at high redshift;
chain galaxies [Cowie et al. 1995 (hereafter C95);
van den Bergh et al. 1996 (hereafter vdB96)].
C95 found 28 candidates of such chain galaxies
in their deep {\it HST} $I$ band (F814W) images of the
two Hawaii deep survey fields, SSA 13 and SSA 22.
vdB96 also identified such chain galaxies
in the Hubble Deep Field (HDF: Williams et al. 1996);
HDF 2--234 (the tadpole or head-tail galaxy) and
HDF 3--531 (the chain galaxy).
The chain galaxies tend to be very straight in morphology; their average
major-to-minor axial ratio is $\sim$ 5.
Another interesting property of
the chain galaxies is that they are bluer on average
than galaxies with similar $I$ magnitudes.
Based on these observational properties,
C95 proposed that the chain galaxies comprise
a new population of forming galaxies
at high redshift ($z \sim$ 0.5 -- 3)
because there is no local counterpart.

However, Dalcanton \& Shectman (1996) proposed a new interpretation
that the chain galaxies are edge-on low surface brightness (LSB) galaxies
at low redshift ($z \sim 0.03$). They first estimated true axial 
ratios of the chain galaxies studied by C95.
They showed that the true axial ratios range between 2 and 14
in most cases but some extreme cases have the ratios higher than 20
if the HST images are deconvolved with the typical point spread
function. These higher axial ratios may favor the edge-on disk 
interpretation. They also showed that the ellipticity distributions
of low-redshift LSB galaxies are consistent with the low-redshift 
chain galaxies being edge-on manifestations of disks using their
database of Las Campanas redshift survey LSB galaxies.

Recently, Bunker et al. (2000) obtained Keck optical spectroscopy of 
the Hot Dog galaxy [= HDF 4-555.1 (Williams et al. 1996) =
C 4-06 (Steidel et al. 1996)] at $z = 2.803$, 
which is one of chain galaxies in the HDF.
They found that any systemic
rotation of this galaxy is fairly small; $<$ 100 km s$^{-1}$.
This suggests strongly that this galaxy is not an edge-on disk
galaxy. Further, the presence of N {\sc v} and He {\sc ii} emission
lines needs massive O stars as their ionization sources at least.
Its blue color also indicates that the stellar populations should
be young ($< 10^8$ yr). This study reinforces the possibility that some 
chain galaxies are really forming galaxies with a linear structure
at high redshift as suggested by C95 (see also Abraham 1997). 
Even if the majority of chain galaxies might be edge-on LSB 
galaxies at low redshift, it seems urgent to investigate 
possible origins of chain galaxies at high redshift such as 
the Hot Dog galaxy.

\section{OBSERVATIONAL PROPERTIES OF CHAIN GALAXIES}

Prior to presenting our new model, we summarize 
important observational properties of chain galaxies
found by C95 and vdB96. A brief summary of the two surveys
is given in Table 1.

\subsection{The C95 Sample}

C95 found 28 candidates of chain galaxies
in the two Hawaii deep survey fields, SSA 13 and SSA 22.
The chain galaxies tend to be very straight and their
major-to-minor axial ratios range from 1.8 to 9.5 with an average of
$4.8 \pm 2.0$ for 21 galaxies brighter than $I = 25$.
Two of them are intermediate-redshift galaxies;
$z = 0.489$ for Chain 0 and $z = 0.505$ for Chain 1.
Another two galaxies are located at $z = 1.36$ (SSA 22--16)
and at $z \sim 2.4$ (SSA 22--24) although the latter
redshift estimate is based on interpreting a break in the
continuum at 4000 \AA ~ as the onset of intergalactic
Ly$\alpha$ forest. 

It is found that the chain galaxies are bluer on average
than galaxies with similar $I$ magnitudes studied by them.
In particular, they are very blue in
$B - K$, suggesting that they are not relatively normal
galaxies in the rest ultraviolet band and that the peculiar
morphologies are not a consequence of the distribution of the
star-forming regions.

\subsection{The vdB96 Sample}

vdB96 classified eleven galaxies either as the chain or as the tadpole
galaxies. Their basic data are given in Table 2. In order to estimate
their linear sizes and absolute magnitudes, we need their redshifts.
However, a spectroscopic redshift is available only for the galaxy 3-379,
the ID number in vdB96 (Steidel et al. 1996): $z = 2.775$.
It is interesting to note that this redshift is very close to 
that of the Hot Dog galaxy, 2.803 (Steidel et al. 1996; Bunker et al. 2000).
For galaxies in the HDF, Fern\'andez-Soto, Lanzetta, \& Yahil (1999) estimated
their photometric redshifts by using the SED (spectral energy distribution)
fitting method\footnote{Hogg et al. (1998) made a comparison between 
various photometric redshift predictions and spectroscopic redshifts
for galaxies in the HDF. They found that the best photometric redshift schemes
predict spectroscopic redshifts with a redshift accuracy of $\Delta z < 0.1$
for more than 68\% of galaxies and with $\Delta z < 0.3$ for  90 -- 100\%.
The photometric redshift for 3-379, $z_{\rm phot}
= 1.72$, appears significantly different from the spectroscopic one, 
$z_{\rm spec} = 2.8$ (see Table 2). However, when we use a certain set of SEDs
generated by using the stellar population synthesis code GISSEL96
(Bruzual \& Charlot 1993), one obtains $z_{\rm phot} \simeq 2.2$.}.
We use these photometric redshifts for the remaining 10
galaxies. Unfortunately, photometric redshifts are not reliably determined
for the two galaxies; 3-367 and 4-627, which are not considered further. 
Then we obtain a sample of the nine chain galaxies. 
We use this sample to investigate the observational properties
of chain galaxies; although the chain galaxies studied by C95 constitute 
the largest sample, their photometric data are less available than
those found in HDF.

The redshifts of the nine galaxies
ranges from 0.5 to 2.8, being similar to the range
estimated for the C95 sample of chain galaxies. 
Adopting a cosmology with the Hubble constant, $H_0 = 100 ~ h$
km s$^{-1}$ Mpc$^{-1}$ where $h=1$ and the density parameter $\Omega_0 = 0.01$,
we estimate the linear size ($d$) and the absolute $V$ magnitude 
($M_V$) for each galaxy. The results are given in Table 2.
We find that $d$ ranges from 1.2 kpc to 3.8 kpc and $M_V$
ranges from $-16.2$ to $-21.5$. Their average values are
$<d> \simeq 2.4 \pm 0.8$ kpc and $<M_V> \simeq -19.0 \pm 1.7$,
respectively. Since any inclination correction has not been made
for the linear sizes, they are lower limits. 
If they are disks viewed edge-on,  the linear sizes appear 
smaller than those of the present-day galaxies.
On the other hand, the absolute
$V$ magnitudes are almost comparable to those of the present-day
galaxies. Frequency distributions of the redshifts, sizes, and 
absolute magnitudes are shown in Figure 1.

The major-to-minor axial ratios of the nine galaxies range from 
1.6 to 5.9 and their average is $3.1 \pm 1.3$, being 
similar to those of the chain galaxies studied by C95.
However, as claimed by Dalcanton \& Shectman (1996),
these axial ratios should be typically larger than 10
if the images are deconvolved with the point spread function.

In summary, the vdB96 chain galaxies 
have the following important observational properties;
$d \sim$ 2.4 kpc, $M_V \sim -19$ (or $L_V \sim 3.4 \times 10^9 L_\odot$),
and $z \sim$ 0.5 -- 3,
all of which are similar to the chain galaxies of C95.
These properties reinforce that a population of chain
galaxies  does actually exist at intermediate and high redshift,
even though edge-on LSBs at low redshifts may share this appearance.

\section{MODEL}

\subsection{Background and Motivation}

A simple explanation for high-$z$ chain galaxies may be that
linear structures form during the collapse of the protogalactic
gas (C95; Bunker et al. 2000).
C95 suggested that when star formation turns on it triggers induced star
formation along the line of maximum density, being analogous to sequential
propagation of OB star associations in molecular clouds.
On the other hand, Bunker et al. (2000) suggested
that star formation could be triggered by collapse along a filament
(see also Abraham 1997).
An alternative idea may be string-induced wakes as noted by C95.
In this paper, we propose another idea that chain galaxies arise from
gaseous filaments which are formed in a large-scale shocked shell
caused either by a superwind or
by interaction between superwinds of forming galaxies.

The starting point of our model is that the superwind activity
occurs in some giant galaxies which were assembled from a number of
gaseous clumps at high redshift within the framework of so-called
cold dark matter scenarios. This assumption implies that initial
starbursts occur in some giant galaxies in their forming phase.
We note below that a couple of lines of evidence for such luminous forming
galaxies have been obtained recently.

\begin{description}

\item 
(1) A number of faint dusty galaxies have been found
by deep surveys at submillimeter (Smail et al. 1997, 1998, 1999, 2000;
Hughes et al. 1998; Barger et al.
1998, 1999a, 1999b; Barger, Cowie, \& Sanders 1999; Barger, Cowie, \&
Richards 2000; Ivison et al. 2000b) and at far infrared
(Kawara et al. 1998; Eales et al. 1999; Puget et al. 1999).
Most of these objects are considered to be located at intermediate and
high redshift (e.g., Ivison et al. 2000b) and thus their bolometric
luminosities often exceed $10^{12} L_\odot$, suggesting a very high
star formation rate, i.e., $\gtrsim 100 M_\odot$ y$^{-1}$.
This implies that they are massive forming galaxies enshrouded by
dusty clouds.

\item
(2) Ly$\alpha$ emitters have been also found at high redshift
$z \gtrsim 2$ (Hu \& McMahon 1996; Cowie \& Hu 1998; Keel et al. 1999;
Steidel et al. 2000). All these are indeed well-defined candidates of
forming galaxies with a high star formation rate.
A remarkable class of such Ly$\alpha$ emitters is called as
^^ ^^ Ly$\alpha$ blobs" (Steidel et al. 2000) which are spatially
extended ionized nebulae. These blobs may be interpreted as
possible evidence for the superwind activity (Taniguchi \& Shioya 2000).

\end{description}
Therefore, it seems reasonable to accept that the superwind activity
works at high redshift.
Although it is unlikely that massive galaxies could be born 
ubiquitously in the early universe, it seems worthwhile investigating
the gravitational instability in shocked shells driven by superwinds
from forming galaxies located in some over-dense regions at high redshift;
such over-dense regions beyond $z = 3$ have 
been found by Steidel et al. (2000) and Ivison et al. (2000a).

It is also noted that we are not simply resurrecting the 
explosion scenarios proposed by Ostriker \& Cowie (1981) and
Ikeuchi (1981). What we assume here is that {\it some} massive galaxies
were assembled from gaseous clumps at high redshift
(i.e., the hierarchical clustering scenario) and then
intense starbursts occurred in them, leading to the superwind
activity in such over-dense regions.

\subsection{Model}

First we give a summary of our model.
Step I: Successive merging of subgalactic gas clumps results in
the formation of a galaxy with a mass of $10^{11-12} M_\odot$
at redshift $z \sim 5$.
Step II: Subsequently, supernova explosions occur inside the galaxy
and then blow out as a galactic wind (or a superwind).
This wind expands into the intergalactic space and then causes a 
large-scale shell with a radius of several hundreds of kpc.
Since this radius may be smaller than the typical separation between
galaxies, interactions of shells may also occur, resulting in
the formation of a large-scale gaseous slab. 
Step III: Since the shell or the slab can be regarded as a gaseous sheet, 
filament-like gravitational instability is expected to occur.   
Step IV: Further gravitational instability occurs in each filament,
leading to intense star formation along the filament. This is the
chain galaxy phase.
Step V: The filament collapses gravitationally into one spheroidal 
system like an elliptical galaxy within one dynamical timescale of
the filament ($\sim 10^8$ yr). A schematic illustration of our model
is shown in Figure 2.

We begin our scenario with the superwind activity expected from 
forming galaxies at high redshift (e.g., Arimoto \& Yoshii 1987).
It is considered that intense star formation (i.e., a starburst) occurs 
at the epoch of galaxy formation in the galaxy center,
producing a galactic wind which lasts for a characteristic
time $t_{\rm GW} (\sim 0.5$ Gyr for an elliptical with a stellar mass 
of $10^{11} M_\odot$: Arimoto \& Yoshii 1987).
Since infalling gas is accreting onto the galaxy
at times $t \geq t_{\rm GW}$, the wind interacts with this gas,
and shocked gaseous shells form
in the outer regions of the galaxy. If the shells are unstable
gravitationally, clumps may be formed within them. 
This explosion-driven formation mechanism has been applied to the 
formation of galaxies (Ostriker \& Cowie 1981; Ikeuchi 1981; Ikeuchi, Tomisaka,
\& Ostriker 1983), dwarf galaxies (Mori, Yoshii, \& Nomoto 1999), 
and globular clusters (Taniguchi, Trentham, \&  Ikeuchi 1999).

A strong observational constraint on such explosion-driven formation scenarios
arises from the very small fluctuation of the cosmic microwave background (CMB)
radiation observed by {\it COBE} (Fixen et al. 1996);
i.e., the dimensionless cosmological distortion parameters are
limited to $\vert y \vert < 1.5 \times 10^{-5}$ and 
$\vert \mu \vert < 9 \times 10^{-5}$ (95\% confidence level)\footnote{
Recently, Battistelli, Fulcoli, \& Macculi (2000) re-estimated
the external calibrator (XCAL) emissivity and then obtained the following
new upper limits;  $\vert y \vert < 3.1 \times 10^{-5}$ and 
$\vert \mu \vert < 4.5 \times 10^{-4}$ (95\% confidence level).
These values are closer to the previous estimates by Mather et al.
(1994). However, in our paper, we adopt the strongest constraints
obtained by Fixen et al. (1996).}.
These limits lead to the fractional energy limit of 
$\Delta U/U = 4 y < 6 \times 10^{-5}$ where $U$ is the energy of the
CMB and $\Delta U$ is the energy converted from other forms.
The energy of the CMB per $L^*$ galaxy is estimated to be
$\sim 10^{63} (1 + z)$ ergs (Wright et al. 1994). 
Therefore, the above fractional energy limit implies that 
the explosion energy per $L^*$ galaxy must be
smaller than $6 \times 10^{58} (1+z)$ ergs.
Since the explosion-driven formation of galaxies proposed by 
Ostriker \& Cowie (1981) and Ikeuchi (1981) violates this limit,
they are rejected by the {\it COBE} observations (Wright et al. 1994). 

Therefore, in our model, we take the fractional energy limit into account.
Assuming that the superwind activity occurs at $z = 5$,
we obtain the upper limit of the energy ejected from an $L^*$ galaxy,
$3.6 \times 10^{59}$ ergs.
This energy corresponds to that released from $3.6 \times 10^8$ supernovae 
because the energy released by a single supernova is $10^{51}{\rm ergs}$.
Thus the total number of supernovae allowed per  $L^*$ galaxy is 
$N_{\rm SN} = 3.6 \times 10^8$. 
Here we assume that the initial starburst occurs obeying 
a Salpeter initial mass function
with the lower-mass and higher-mass cutoffs, $M_l = 0.1 ~ M_{\odot}$ and
$M_u = 100 ~ M_{\odot}$, respectively.  It is known that 
massive stars  with a mass of $m > 8 M_{\odot}$
are progenitors of type II supernovae. In the above star formation,
the number of such high mass stars per unit gas mass ($1 M_{\odot}$)
is estimated to be $\simeq$ 0.007.
Therefore, the mass of a galaxy which produce $N_{\rm SN} = 3.6 \times 10^8$ 
is estimated as $M_{\rm gal} \sim N_{\rm SN}/0.007 \sim 5 \times 10^{10} M_\odot$,
being roughly comparable to that of an  $L^*$ galaxy.

Then we consider the physical conditions of the shell formed by
the superwind from such an $L^*$ galaxy
following Ostriker \& Cowie (1981).
In the radiative cooling epoch, the radius of the cooled shell is

\begin{equation}
R_{\rm cool} = 3 E_{61}^{0.3} \left[ h^2 \Omega_0 (1+z)^3 \right]^{-0.4} \; {\rm Mpc},
\label{eqn:ocr}
\end{equation}
and the gas mass swept out to the shell is

\begin{equation}
M_{\rm cool} = 2 \times 10^{13} E_{61}^{0.88} \left[ h^2 \Omega_0 (1+z)^3
\right]^{-0.21} M_{\odot}, \label{eqn:ocm}
\end{equation}
together with the cooling time,

\begin{equation}
t_{\rm cool} = 6 E_{61}^{0.2} \left[ h^2 \Omega_0 (1+z)^3 \right]^{-0.5} \; {\rm Gyr}.
\end{equation}
where $E_{61}$ is the explosion energy in units of $10^{61}$ ergs.
If we assume that a shell is formed from this swept-out gas, 
its surface mass density  is

\begin{equation}
\Sigma_{\rm cool}  =  \frac{M_{\rm cool}}{4 \pi R_{\rm cool}^2} 
  = 3.71 \times 10^{-5} E_{61}^{0.28}h^{1.18}\Omega_0^{0.59}(1+z)^{1.77}
{\rm g \; cm^{-2}}.
\end{equation}
Adopting $\Omega_0 = 0.01$, $h = 1$, $z = 5$, and $E_{61} = 0.036$,
we obtain $R_{\rm cool} = 0.81$ Mpc,
$M_{\rm cool} = 9.1 \times 10^{11}M_{\odot}$, and 
$\Sigma_{\rm cool}=2.3 \times 10^{-5}$ g cm$^{-2}$ for the shell
formed by the superwind from the $L^*$ galaxy.

Since the cooling time is estimated as $\simeq$ 2.1 Gyr,
the shell will form at redshift $z_{\rm shell} \sim 1.6$
given the above cosmological parameters.
Since the number density
of galaxies in the present-day universe is $n_{\rm G}(0) \sim 0.01$ Mpc$^{-3}$,
$n_{\rm G}(z)$ is estimated as $n_{\rm G}(0) (1+z)^3$.
Given the adopted cosmology, we obtain 
$n_{\rm G}(1.6) \simeq$ 0.18 Mpc$^{-3}$, giving 
an average separation, $L_{\rm G} = n_{\rm G}^{-1/3}
(4 \pi / 3)^{-1/3} \simeq$ 1.1 Mpc.
Although this is slightly larger than $R_{\rm cool}$,
it is suggested that such superwinds will overlap in some cases
because we assume that the superwinds occur from giant galaxies
in over-dense regions of the universe.
In either case, a large-scale shocked shell with a radius of $\sim$ 1 Mpc
can be formed in this model.
The physical parameters of the shell are summarized in Table 3
for $\Omega_0$ = 0.01, 0.1, and 1 and $h$ = 0.5, 0.75, and 1.

Then we investigate what happens in such a shell. 
Since the radius of the shell is so large, 
the shell can be approximated on sufficiently small scales
(i.e., $\sim$ 10 kpc) by a planer, self-gravitating layer,
or  sheet (e.g., Elmegreen \& Elmegreen 1978).
It is known that gravitational instability in an isothermal 
self-gravitating sheet results in the formation of filaments
(Miyama, Narita, \& Hayashi 1987a, 1987b; Curry 2000 and references
therein). According to Curry (2000),
the critical wavelength of the filamentary instability is defined as 

\begin{equation}
\lambda_{\rm cr} = c_{\rm s} [2 \pi / (G \rho_{\rm shell})]^{1/2}
\end{equation}
where $c_{\rm s}$ is the sound speed, $\rho_{\rm shell}$ is the mass 
density of a sheet and $G$ is the gravitational constant.
If we assume that $\rho_{\rm shell} \sim 10 \rho_{\rm cr}(z)$
where $\rho_{\rm cr}(z) = \rho_{\rm cr}(0) (1+z)^3$, we obtain
$\rho_{\rm shell} \simeq 3.34 \times 10^{-27}$ g cm$^{-3}$ at $z=1.6$.
Since it is expected that the sound velocity is $c_{\rm s} \sim$ 10 km s$^{-1}$
(e.g., Ostriker \& Cowie 1981), we obtain $\lambda_{\rm cr} \simeq$ 54 kpc.
Using the mass density together with the surface mass density of the shell,
we obtain  the thickness of the shell (sheet),
$\Delta R = \Sigma_{\rm cool}/\rho \sim 2.2$ kpc. 
Thus an initial major-to-minor axial ratio is $\sim$ 25,
being consistent with the axial ratios estimated for the 
chain galaxies (see Dalcanton \& Shectman 1996).
According to Miyama et al. (1987a, 1987b), it is expected that
this filament experiences the beads instability and then collapses
along the major axis of the filament, forming an elliptical-like
galaxy. Even if a cigar-like, prolate galaxy could be formed from
the filament, it would be dynamically unstable and thus evolve
to a rounder elliptical galaxy (Merritt \& Hernquist 1991).

The mass density of the shell leads to the dynamical 
timescale of the sheet, $\tau_{\rm dyn} \sim (G \rho_{\rm shell})^{-1/2}
\simeq 2.1 \times 10^9$ yr. Since this timescale is comparable to
the cooling time, it is considered that
chain galaxies can be formed at $z \simeq 1.6$. However, we note that
chain galaxies could be born preferentially in higher-density
regions in the shocked shell and thus the formation epoch may be earlier than
the above redshift. In order to explain the formation of 
chain galaxies at $z \sim 3$, one may assume that the epoch of the
superwind activity occurs at $z > 5$.

Now let us estimate the mass of the filament considered above.
The filament could grow in mass by accumulating gas clouds 
within a roughly regular square region with a length of $\lambda_{\rm cr}$
and a thickness of $\Delta R$. Therefore, we estimate the mass of the
filament, $M_{\rm filament} \sim \lambda_{\rm cr}^2 ~ \Delta R ~ \rho
\sim 3 \times 10^8 ~ M_\odot$. Since a typical luminosity-to-mass ratio of
O stars is $\sim 10^3$, this filament becomes as luminous as 
$\sim 10^{11} L_\odot$.
Note that this luminosity is achieved
if a burst of massive-star formation occurs in it
with very high star-formation efficiency (approximately 100\%).
The reason for such high efficiency is that a filament-like structure
is inherently unstable and will disrupt on a time scale comparable to
the transverse crossing time; $\tau_{\rm cross} \simeq 0.72 \times 10^8 
(d/ {\rm 2.4~kpc}) (v/{\rm 10 ~ km ~s^{-1}})^{-1}$  yr (C95).
The above luminosity is roughly comparable to the observed 
luminosities of the chain galaxies although they show a large scatter
from $\sim 7 \times 10^{8} L_\odot$ to $\sim 4 \times 10^{10} L_\odot$ 
in $L_V$ (see Table 2); this scatter may be attributed to
the density fluctuation in the shocked shell.
After the intense star formation ceases ($\sim 10^8$ years after the
formation; Abraham 1997), this filament 
rapidly fades and then evolves dynamically to a dwarf 
elliptical-like galaxy with $\lesssim 10^8 L_\odot$. This may explain
why we cannot find remnants of the chain galaxies.  

Finally we estimate how many chain galaxies are formed in the shocked shell.
Since the mass of the shocked shell is $M_{\rm cool}
\simeq 9.1 \times 10^{11} M_\odot$ and that of a chain galaxy is
$M_{\rm chain} \sim 3 \times 10^8 M_\odot$, the maximum number of
dwarf galaxies is $N_{\rm chain} \sim 3000 \eta_{\rm chain, 100}$
where $\eta_{\rm chain, 100}$ is the formation efficiency of chain galaxies
in the shell in units of 100\%. Since the radius of the shocked 
shell is $R_{\rm cool} \simeq$ 0.81 Mpc, the volume number density of
chain galaxies is estimated as $n_{\rm chain} \sim 1350 \eta_{\rm chain, 100}$
Mpc$^{-3}$.
However, it is unlikely that galaxies were formed with such the
very high efficiency because the chain galaxy formation could occur
preferentially in higher-density regions in the shell.
If we adopt $\eta_{\rm chain, 100} = 0.1$ (i.e., 10\% efficiency;
see section 4.2),
we obtain $n_{\rm chain} \sim$ 135 Mpc$^{-3}$. 

Given $\eta_{\rm chain, 100} = 0.1$, 300 chain galaxies could be
formed in one shocked shell. 
The shocked shell radius, $R_{\rm cool} \simeq$ 0.81 Mpc,
corresponds to an angular size of 2.5 arcmin at $z \sim 1$.
This leads to a nominal surface number density of chain galaxies
formed in one shocked shell
$\sigma_{\rm chain} \simeq 300 / (\pi 2.5^2) \simeq 15$ arcmin$^{-2}$.
However, since the lifetime of the chain galaxy phase
($\tau_{\rm chain} \sim 10^8$ yr)
is much shorter than the Hubble time ($\tau_{\rm Hubble} \sim 10^{10}$ yr),
the surface number density of chain galaxies made in one shocked shell
is statistically estimated as $\sigma_{\rm chain, stat} \sim
\sigma_{\rm chain} \tau_{\rm chain} \tau_{\rm Hubble}^{-1}
\sim 0.15$ arcmin$^{-2}$. As summarized in Table 1, the observed 
surface density of the chain galaxies is a few chain galaxies
arcmin$^{-2}$. Comparing this with the above estimate, we suggest that
ten shocked shells at different redshifts
(e.g., $z \sim$ 0.5 -- 3) are included in the sky areas of the deep survey fields.
This also implies that only a small number of galaxies could experience
the superwind activity leading to the formation of a shocked shell.

Our model also predicts that several chain galaxies can be found at 
nearly the same redshift because they are associated with one shocked shell.
It is again interesting to note that the redshift of the Hot Dog galaxy
($z = 2.803$) is very close to the chain galaxy 3-379 in vdB96 ($z = 2.775$).
Among the eight galaxies with their photometric redshifts (Table 2),
the following two pairs have the same redshifts; 1) $z_{\rm phot} = 0.92$
for 2-234 and 3-621, and 2) $z_{\rm phot} = 1.60$ for 4-241 and 3-531.
Although most chain galaxies are unfortunately very faint
[$I_{\rm 814}$(AB) $\gtrsim 23$],
it should be possible to perform optical/NIR spectroscopy of the chain
galaxies found by C95 and vdB96 in the near future.

\section{DISCUSSION}

\subsection{How Can Parent Galaxies of Chains Be Seen ?}

Our model may be one of possible mechanisms to form chain galaxies
at high redshift. Our model predict that 
there should be massive parent galaxies located within 1 Mpc 
which produced the shells during a burst of star
formation at much higher redshift. They can be seen as
passively-evolving galaxies at nearly the same redshift
as that of certain chain galaxies.
Using photometric redshifts of the HDF galaxies provided by
Fern\'andez-Soto et al. (1999), we try to search for such parent galaxies.

The most interesting chain galaxy in the vdB96 sample is the chain
galaxy 3-367 at $z = 2.775$ because the Hot Dog galaxy is also located
at nearly the same redshift. However, since Fern\'andez-Soto et al. (1999)
give its photometric redshift $z_{\rm phot} = 1.72$ (see Table 2),
it seems difficult to search for its parent galaxy candidates.
Then, as a trial, we try to search for parent galaxy candidates
for the following two pairs of chain galaxies; 1) vdB96 2-234 and 3-621
at $z_{\rm phot} = 0.92$, and 2 ) vdB96 4-241 and 3-531 at  $z_{\rm phot} = 1.60$.

Since the bin size of photometric redshift studied by Fern\'andez-Soto et al. 
(1999) is $\delta z = 0.02$, we try to find galaxies with  $z_{\rm phot} = 
0.92$ or 1.60. 
We have found 40 galaxies at $z = 0.92$ and 35 ones at $z = 1.60$ in the HDF. 
However, parent galaxies should be passively-evolving ones and thus we reject
galaxies with blue SEDs such as those of irregular galaxies. 
Then we find three possible candidates whose SEDs are relatively red 
at $z = 0.92$ and only one at $z = 1.60$.
The results are summarized in Table 4.
Among the three galaxies at $z = 0.92$, two galaxies (Nos. 531 and 536) 
have lower luminosities than $L^*$. Therefore, the remaining galaxy (No. 466)
may be the most probable candidate although its
projected separation is 0.58 Mpc from vdB96 3-621 and 0.73 Mpc from vdB96 2-234.
As for the candidate galaxy at $z = 1.60$, its luminosity is higher than
$L^*$ and the projected separation is 0.26 Mpc from vdB96 3-531 and 0.25 Mpc from
vdB96 4-241. Therefore, this seems a promising candidate of the parent galaxy.

Here it is again  noted that 
the accuracy of photometric redshifts is not high enough to identify
the parent galaxies conclusively.
In order to find firmer candidates, we will need the spectroscopic 
redshift for a large number of galaxies in the HDF.

\subsection{Connection to the Formation of Dwarf Galaxies}

As described in section 3, it is expected
that the chain galaxies formed in our model will evolve to dwarf galaxies
because their masses are of the order of $\sim 10^8 M_\odot$.
Therefore, our scenario may also be related to the formation of
dwarf galaxies.

First, we estimate the maximum number of dwarf galaxies
formed in the shocked shell.
Since we assume that dwarf galaxies are the descendents of chain galaxies,
the maximum number is the same as that of the chain galaxies; i.e.,
$N_{\rm dwarf} = N_{\rm chain} \sim 3000 \eta_{\rm chain, 100}$. 
Thus, the volume number density of dwarf galaxies is estimated as
$n_{\rm dwarf} = n_{\rm chain} \sim 1350 \eta_{\rm chain, 100}$ Mpc$^{-3}$. 

Recently, Gnedin (2000) investigated an average star formation law 
during the first 3 Gyr using the observational data of Local Group
dwarf galaxies. His analysis showed that the data are consistent with
the orthodox Schmidt law with a star formation efficiency of about 4\%.
Since our model is applicable to galaxies in over-dense regions,
the formation efficiency may be higher than the above estimate.
Therefore, it seems reasonable to assume that the formation of chain galaxies
in the shocked shell is roughly 10\%, as adopted in section 3.
Then,  the volume number density of dwarf galaxies
is down to $n_{\rm dwarf} \sim$ 135 Mpc$^{-3}$. This is 
significantly lower than 
the observed number density of dwarf (elliptical) galaxies in the Virgo
cluster (i.e., an over-dense region in the local universe) 
is $\sim$ 300 Mpc$^{-3}$ (Binggeli, Sandage, \& Tammann 1988).
Although chain galaxies could contribute to the formation of 
dwarf galaxies, 
the majority of the dwarf galaxies in the Virgo cluster may be
formed from other mechanisms; e.g., the gravitational collapse of
less-massive gas clumps or the tidally-induced formation (e.g.,
Okazaki \& Taniguchi 2000 and references therein).

\subsection{Connection to Lyman Limit Systems and Damped Ly$\alpha$ Absorbers}

The shocked gaseous shell formed by the superwind activity 
may be important absorbing material in the intergalactic space.
Indeed, the surface mass density of the shocked shell,
$\Sigma_{\rm cool} \simeq 2.3 \times 10^{-5}$ g cm$^{-2}$, 
corresponds to the H {\sc i} column density, 
$N_{\rm HI} \simeq 1.4 \times 10^{19}$ cm$^{-2}$.
Since the temperature of IGM may be $\sim 10^4$ K at most,
the ionization correction seems modest. 
Therefore, shocked shells formed by forming massive galaxies
may cause some Lyman limit systems ($N_{\rm HI} \gtrsim 10^{17}$ cm$^{-2}$:
e.g., Steidel 1990) found in high-$z$ quasar spectra. Furthermore, 
if we see such shells from a highly inclined angle (e.g., $\theta_{\rm view}
\gtrsim  80^\circ$) or their tangential sections, 
the H {\sc i} column density exceeds $10^{20}$ cm$^{-2}$,
causing damped Ly$\alpha$ systems (e.g., Peebles 1993; Wolfe et al. 1995).
Or, more likely, the damped Ly$\alpha$ systems may arise
in the last stages of gravitational collapse of the filaments
just as star formation commences.
Small cloudlets breaking into intergalactic medium (IGM)
may be observed as Ly$\alpha$ forests
with $N_{\rm HI} \lesssim 10^{15}$ cm$^{-2}$. 

The shock velocity is given by

\begin{equation}
v_{\rm cool} = 200 E_{61}^{0.05} \left[ h^2 \Omega_0 (1+z)^3 \right]^{0.11} \;
{\rm km~ s^{-1}}
\end{equation}
(Ostriker \& Cowie 1981). Adopting $E_{61} = 0.036$, $h = 1$, $\Omega_0 = 0.01$,
and $z = 1$, we obtain $v_{\rm cool} \simeq 130$ km s$^{-1}$. 
Therefore, the absorption
feature caused by one shocked shell may have the line width of $\lesssim$ 
260 km s$^{-1}$. However, the 
velocity widths observed for most high-column-density absorption systems
(determined from unsaturated low-ion species) are much less than 260 km s$^{-1}$.
It is also noted that our model is applicable to the case that an absorbing
galaxy is located significantly far from the quasar; i. e., the projected 
distance up to several hundreds kpc is acceptable. 

In our model, some giant galaxies {\it in over-dense regions} in the universe
could experience initial starbursts and then show the superwind activity.
Therefore, shocked shells may not be formed ubiquitously in high-$z$ universe.
As mentioned in section 3, the observed surface density of chain galaxies
suggests that there are only several shocked shells at most in the redshift
interval between 0.5 and 3. Therefore, our model predicts a
few high-$z$ Lyman limit or damped Ly$\alpha$ systems per unit
redshift along a 1D sight-line."
Stengler-Larrea et al. (1995) showed that the number density of 
Lyman limit systems as a function of redshift is described by
$N(z) \simeq  N_0 (1+z)^\gamma$ where $N_0 = 0.25$ and $\gamma = 1.5$.
This gives $N(3) \simeq 2$, being consistent with our prediction.

Another interesting issue related to our model is 
the metal enrichment in the IGM.
One important consequence of our model is that such H {\sc i} 
absorber must be polluted chemically because the superwind
contains a lot of heavy elements. 
It has been argued that supernova explosions lead to the chemical
enrichment of IGM at high redshift (e.g., Ostriker \& Gnedin 1996;
Miralda-Escud\'e \& Rees 1997; Aguirre et al. 2000).
We estimate the metal enrichment due to the superwind.
The mass of metal ejected from a star is $m_Z = \epsilon_Z m_*$
where $\epsilon_Z$ is the mass fraction of metal with respect to
the stellar mass. Thus the total metal enrichment due to the
superwind can be estimated as

\begin{equation}
\Delta Z  ={{\epsilon_Z m_* N_{\rm SN}} \over {4 \pi R_{\rm cool}^2 \Delta R
\rho_{\rm shell}}} \simeq 4.2 \times 10^{-3} \epsilon_Z
\end{equation}
where we adopt $m_* = 10 M_\odot$ as the typical mass of progenitors of 
type II supernovae.
We then obtain $\Delta Z  \simeq  4.2 \times 10^{-4}$ if $\epsilon_Z = 0.1$.
Since  the metal abundance observed in
Ly$\alpha$ forests (e.g., Pettini et al. 1997; Songaila \&
Cowie 1996; Lu et al. 1998) is
of the order of 0.01 $Z_\odot$ where $Z_\odot$ is the solar metal
abundance in mass ($Z_\odot = 0.02$); i.e., $Z_{\rm IGM}
\simeq 2 \times 10^{-4}$. Therefore, the metal
ejection as a result of the superwind from a forming massive galaxy
may be responsible for the observed metal abundance in the IGM
at high redshift.

\section{CONCLUDING REMARKS}

The chain galaxies found in the recent optical deep surveys 
have been investigated in terms of the gravitational instability
of shocked shells formed by a superwind or
by interaction between superwinds of forming galaxies at high redshift.
Our new model appears consistent with the observational properties of
the chain galaxies. Since the chain galaxies are expected to 
evolve to dwarf galaxies with a mass of $\sim 10^8 M_\odot$,
our model explains why there is no local counterpart.

We have also discussed that shocked shells may be responsible 
for some Lyman limit systems and damped Ly$\alpha$ systems 
observed in quasar spectra because their H {\sc i} column
densities exceed $N_{\rm HI} \gtrsim 10^{17}$ cm$^{-2}$. 
The metal enrichment up to $Z \sim 0.01 Z_\odot$ can be 
achieved by the pollution from supernova ejecta.
In order to investigate this possibility further, we need
very high-resolution, emission-line imaging of some 
counterparts of damped Ly$\alpha$ systems at high redshift. 

The size of such shocked shells is too large (i.e., $\sim$ 1 Mpc)
to be detected if the sky coverage of deep surveys is not so large. 
In order to find firm observational evidence for such shocked shells,
wide-field, deep surveys will be also very important in future. 
The SUPRIME-Cam (the Subaru prime focus camera; e.g., 
http://SubaruTelescope.org/Introduction/instrument.html)
seems to be the most efficient tool to perform such surveys.

\vspace {0.5cm}

We would like to thank S. Ikeuchi, S. Miyama, and M. Hattori for
giving us useful scientific input and S. Okamura for useful discussion 
on the capability of SUPRIME-Cam on the Subaru telescope.
We would also thank an anonymous referee for his/her very useful 
comments and suggestions which improved this paper significantly.
YS is a JSPS fellow. This work was financially supported in part by
the Ministry of Education, Science, and Culture
(Nos. 10044052, and 10304013).

\newpage

\newpage

\begin{deluxetable}{lcc}
\tablecaption{%
A summary of the surveys by C95 and vdB96
}
\tablehead{
   \colhead{} & 
   \colhead{C95} &
   \colhead{vdB96}
}
\startdata
Field & SSA 13 \& SSA 22 & HDF \nl
Area  & 10.7 arcmin$^2$ & 5.3  arcmin$^2$ \nl
$N_{\rm chain}(I_{814} \leq 25)$\tablenotemark{a} & 24 & 11 \nl
$N_{\rm chain}(I_{814} > 25)$\tablenotemark{b} & 4 & \nodata \nl
$\sigma_{\rm chain}(I_{814} \leq 25)$\tablenotemark{c} & 2.2 arcmin$^{-2}$ &
        2.1 arcmin$^{-2}$ \nl
$\sigma_{\rm chain}(I_{814} >  25)$\tablenotemark{d} & 0.4 arcmin$^{-2}$ & \nodata \nl
$<a/b>$\tablenotemark{e} & $4.8 \pm 2.1$ & $3.1 \pm 1.3$ \nl
\enddata
\tablenotetext{a}{The observed number of chain galaxies with $I_{814} \leq 25$.}
\tablenotetext{b}{The observed number of chain galaxies with $I_{814} > 25$.}
\tablenotetext{c}{The surface density of chain galaxies with $I_{814} \leq 25$.}
\tablenotetext{d}{The surface density of chain galaxies with $I_{814} > 25$.}
\tablenotetext{e}{The average value of apparent major-to-minor axial ratio.}
\end{deluxetable}

\newpage

\begin{deluxetable}{llccccccccc}
\tablecaption{%
Observational properties of the vdB96 chain galaxies
}
\tablehead{
   \colhead{ID\tablenotemark{a}} &
   \colhead{ID\tablenotemark{b}} &
   \colhead{$r_1$\tablenotemark{c}} &
   \colhead{$d$\tablenotemark{d}} &
   \colhead{$a/b$\tablenotemark{e}} &
   \colhead{AB(814)\tablenotemark{f}} &
   \colhead{$z_{\rm spec}$\tablenotemark{g}} &
   \colhead{$z_{\rm phot}$\tablenotemark{h}} &
   \colhead{$M_V$\tablenotemark{i}} &
   \colhead{$L_V$\tablenotemark{j}} \nl
   & & \colhead{($^{\prime \prime}$)} & \colhead{(kpc)} & 
   & \colhead{(mag)} & & & \colhead{(mag)} & \colhead{($L_{\odot}$)}
}
\startdata
4-627 & 424& 0.16& \nodata & 1.12& 23.28& \nodata & \nodata & \nodata & \nodata \nl
3-367 & 363& 0.24& \nodata & 2.63& 24.43& \nodata & \nodata & \nodata & \nodata \nl
2-234 & 1027& 0.55& 2.90 & 2.63& 23.88& \nodata & 0.920 & $-18.5$& $2.1\times 10^9$ \nl
4-375 & 81& 0.37& 1.69 & 3.33& 23.94& \nodata & 0.640 & $-17.3$& $7.0\times 10^8$ \nl
4-241 & 679& 0.55& 3.39 & 2.63& 23.94& \nodata & 1.600 & $-20.5$& $1.4\times 10^{10}$ \nl
3-243 & 496& 0.45& 1.85 & 2.50& 24.51& \nodata & 0.520 & $-16.2$& $2.6\times 10^8$ \nl
3-621 & 16& 0.49& 2.59 & 5.26& 24.93& \nodata & 0.920 & $-17.4$& $7.6\times 10^8$ \nl
3-379 & 334& 0.34& 2.27 & 2.32& 24.71& 2.775 & 1.720 & $-21.5$& $3.5\times 10^{10}$ \nl
2-526 & 836& 0.25& 1.63 & 2.38& 25.27& \nodata & 2.240 & $-20.2$& $1.0\times 10^{10}$ \nl
3-448 & 276& 0.20& 1.24 & 2.88& 25.35& \nodata & 1.640 & $-19.4$& $5.1\times 10^{9}$ \nl
3-531 & 120& 0.61& 3.75 & 5.88& 24.27& \nodata & 1.600 & $-20.2$& $1.0\times 10^{10}$ \nl
\enddata
\tablenotetext{a}{ID number given in vdB96}
\tablenotetext{b}{ID number given in Fern\'andez-Soto et al. (1999)}
\tablenotetext{c}{Apparent size along the major axis}
\tablenotetext{d}{Linear size along the major axis}
\tablenotetext{e}{Apparent major-to-minor axial ratio}
\tablenotetext{f}{AB magnitude at $I$ band}
\tablenotetext{g}{Spectroscopic redshift}
\tablenotetext{h}{Photometric redshift taken from
                  Fern\'andez-Soto et al. (1999)}
\tablenotetext{i}{Absolute $V$ magnitude}
\tablenotetext{j}{$V$ band luminosity}
\end{deluxetable}

\newpage

\begin{deluxetable}{ccccccc}
\tablecaption{%
Parameters of the shell
}
\tablehead{
   \colhead{$\Omega_0$} &
   \colhead{$h$} &
   \colhead{$t_{\rm cool}$} &
   \colhead{$R_{\rm cool}$} &
   \colhead{$M_{\rm cool}$} &
   \colhead{$\Sigma_{\rm cool}$} &
   \colhead{$\Delta R$} \nl
   &  & \colhead{(Gyr)} & \colhead{(Mpc)} & \colhead{($M_{\odot}$)} &
   \colhead{(g cm$^{-2}$)} & \colhead{(kpc)}
}
\startdata
0.01& 1& 2.1& 0.81& $9.1 \times 10^{11}$& $2.3 \times 10^{-5}$& 2.2  \nl
0.1 & 1&0.66& 0.32& $5.6 \times 10^{11}$& $9.0 \times 10^{-5}$& 8.7  \nl
1.0 & 1&0.21& 0.13&$3.5 \times 10^{11}$& $3.5 \times 10^{-4}$& 34   \nl
0.01&0.75& 2.8&1.0& $1.0 \times 10^{12}$& $1.6 \times 10^{-5}$& 1.6  \nl
0.1 &0.75&0.89&0.41&$6.4 \times 10^{11}$& $6.4 \times 10^{-5}$& 6.2  \nl
1.0 &0.75&0.28&0.16&$3.9 \times 10^{11}$& $2.5 \times 10^{-4}$& 24   \nl
0.01&0.5&4.2& 1.4& $1.2 \times 10^{12}$& $1.0 \times 10^{-5}$& 0.99 \nl
0.1 &0.5&1.3& 0.56& $7.5 \times 10^{11}$& $4.0 \times 10^{-5}$& 3.8  \nl
1.0 &0.5&0.42&0.22&$4.6 \times 10^{11}$& $1.5 \times 10^{-4}$& 15   \nl
\enddata
\end{deluxetable}


\newpage
\begin{deluxetable}{cccccccc}
\tablecaption{%
Parent galaxy candidates
}
\tablehead{
   \colhead{ID\tablenotemark{a}} &
   \colhead{SED\tablenotemark{b}} &
   \colhead{$M_V$\tablenotemark{c}} &
   \colhead{$L_V$\tablenotemark{d}} &
   \multicolumn{2}{c}{$D_1$\tablenotemark{e}} &
   \multicolumn{2}{c}{$D_2$\tablenotemark{f}} \nl
   \cline{5-6}
   \cline{7-8}
   & & \colhead{(mag)} & \colhead{($L_{\odot}$)} &
   \colhead{($^{\prime \prime}$)} & \colhead{(Mpc)} &
   \colhead{($^{\prime \prime}$)} & \colhead{(Mpc)}
}
\startdata
\multicolumn{8}{c}{z=0.92} \nl
\hline
466 & Sbc & $-20.7$ & $1.67 \times 10^{10}$& 109& 0.58& 138& 0.73 \nl
531 & M31 & $-18.8$ & $2.73 \times 10^9$   &  62& 0.33&  90& 0.48 \nl
567 & M31 & $-17.8$ & $1.14 \times 10^9$   &  70& 0.37&  91& 0.48 \nl
\hline
\multicolumn{8}{c}{z=1.60} \nl
\hline
306 & Sbc & $-21.0$ & $2.05 \times 10^{10}$&  43& 0.26&  41& 0.25 \nl
\enddata
\tablenotetext{a}{ID number given in Fern\'andez-Soto et al. (1999)}
\tablenotetext{b}{Best-fit SED type in Fern\'andez-Soto et al. (1999)}
\tablenotetext{c}{Absolute $V$ magnitude}
\tablenotetext{d}{$V$ band luminosity}
\tablenotetext{e}{Distance from 3-621 (z=0.92) or 3-531 (z=1.60)}
\tablenotetext{f}{Distance from 2-234 (z=0.92) or 4-241 (z=1.60)}
\end{deluxetable}

\newpage

\begin{figure*}
\epsscale{0.6}
\plotone{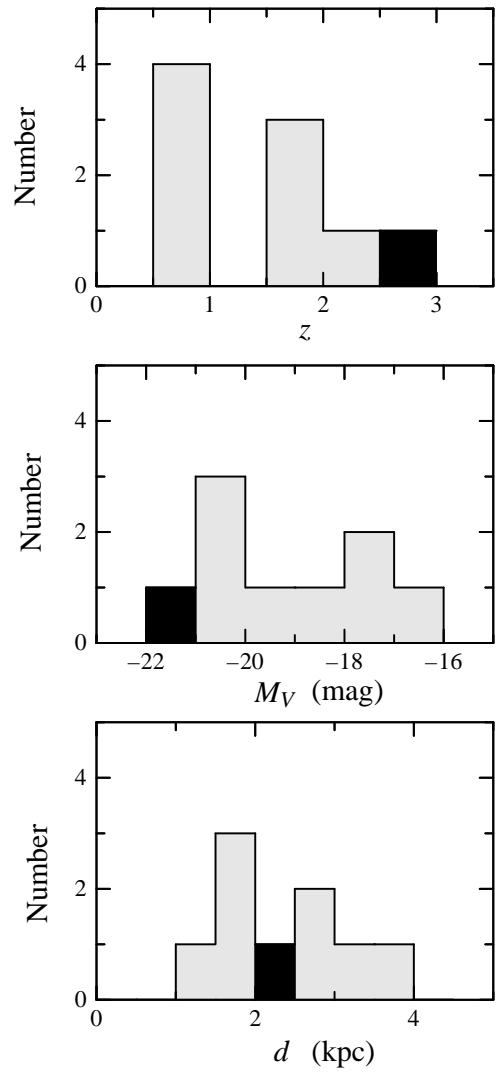}
\caption{
Frequency distributions of the redshifts, sizes, and
absolute magnitudes of the nine chain galaxies in vdB96.
The galaxy 3-379 is shown by the filled square in each
panel because it has the spectroscopic redshift.
\label{fig1}}
\end{figure*}

\begin{figure*}
\epsscale{0.6}
\plotone{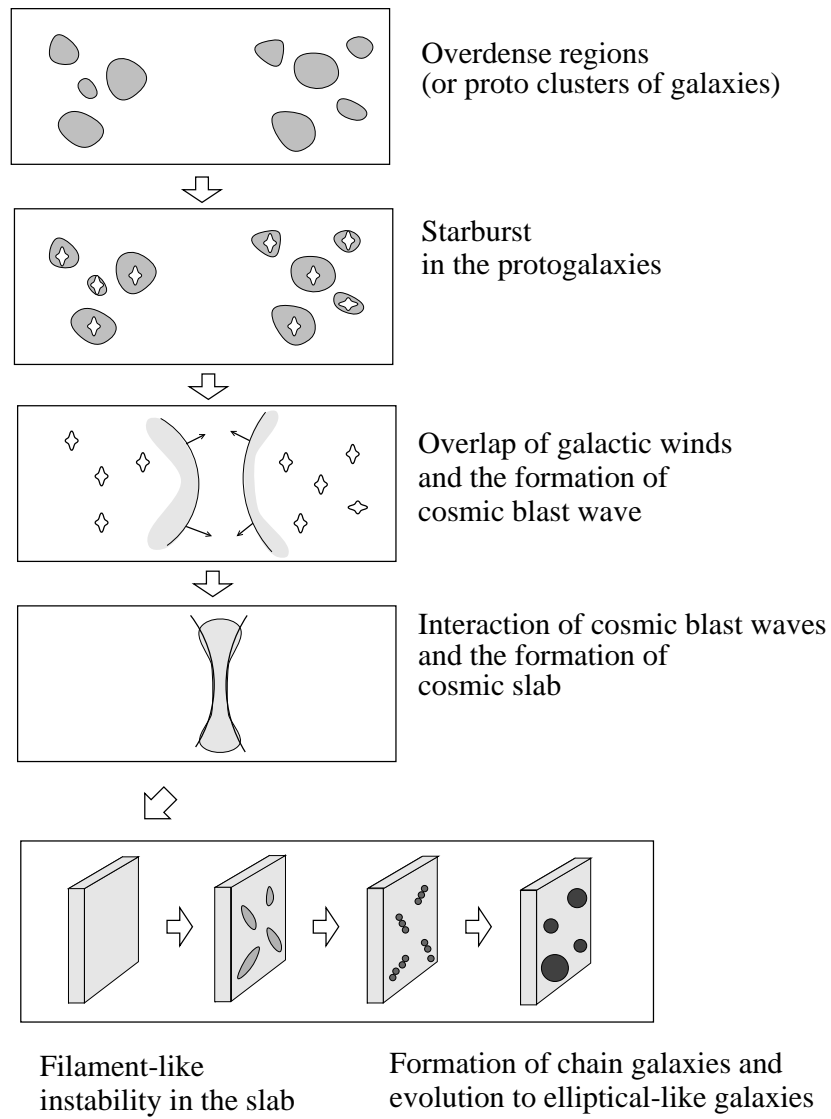}
\caption{
A schematic illustration of the proposed formation mechanism
of chain galaxies.
\label{fig2}}
\end{figure*}

\end{document}